\documentclass[reprint,amsmath,amssymb,aps, twocolumn]{revtex4-2}

\usepackage{hyperref}
\usepackage{comment}
\hypersetup{colorlinks,urlcolor=blue}
\usepackage{graphicx}
\usepackage[percent]{overpic}
\usepackage{dcolumn}
\usepackage{bm}
\usepackage{ulem}
\usepackage{physics}
\usepackage{lipsum}
\usepackage{appendix}
\usepackage[T1]{fontenc}

\newcommand{\red}[1]{\textcolor{red}{#1}}
\newcommand{\ham}{\mathcal{H}}
\newcommand{\vect}[1]{\boldsymbol{#1}}
\newcommand{\expv}[1]{\langle{#1}\rangle}
\newcommand{\ensavg}[1]{\overline{#1}}
\newcommand{\timeft}[1]{\widetilde{#1}}

\begin{document}
\preprint{APS/123-QED}
\title{Dynamics of spin helices in the diluted one-dimensional $XX$ model}
\author{Darren Pereira}
\author{Erich J. Mueller}
\affiliation{Laboratory of Atomic and Solid State Physics, Cornell University, Ithaca, New York 14853, USA}

\begin{abstract}

Motivated by discrepancies between recent cold atom experiments and the associated theory, we explore the effect of immobile holes on the quantum dynamics of $x$--$z$ spin helices in the one-dimensional $XX$ model. We calculate the exact spin dynamics by mapping onto a system of non-interacting fermions, averaging over the distribution of holes.  
At small hole densities we find that the helical spin pattern decays exponentially, with a pitch dependence that agrees with the experiments.
At large hole densities we instead find persistent oscillations.  
While our analytic approach does not generalize to the $XXZ$ model with arbitrary anisotropies, we validate a matrix product state technique which might be used to model the experiments in those settings.
\end{abstract}

\date{\today}
\maketitle

\section{Introduction}
\label{sec:Intro}

In a recent experiment \cite{JepsenNature2020}, Jepsen et al. used a gas of $^7$Li atoms trapped in a one-dimensional (1D) optical lattice as an analog simulator of the 1D anisotropic Heisenberg model, one of the most important spin models from condensed matter physics \cite{Heisenberg1926, LSMAnnals1961, HaldanePRL1983, AKLTPRL1987, AffleckJPCM1989, TasakiPRL1991, ProsenNatComm2017, ProsenPRL2019, VasseurPRL2019, RichardsAnnuRevMS1974, BoseCP2007, HeinrichRMP2019, BertiniRMP2021, GopalakrishnanAnnuRevCMP2023}. In certain limits their experiment showed behavior which was qualitatively different from the exact solution of that model \cite{PereiraSpinHelix2022, VasseurHelixPRB2023}.  Here we show that 
adding a realistic density of
immobile holes to the model 
removes the discrepancy.

In the experiment bosonic lithium atoms were loaded into an optical lattice and confined with an additional harmonic trap.  Due to the deep optical lattice, the atoms formed a Mott insulator, and in the majority of the cloud there was a single particle per site.
Superexchange, from the virtual hopping between neighboring sites, led to an effective $XXZ$ (or anisotropic Heisenberg) model
$\ham=\sum_j J_x (S^x_j S^x_{j+1}+S^y_j S^y_{j+1})+J_z S^z_j S^z_{j+1}$.  Here, $J_x$ and $J_z$ parameterize the nearest-neighbor interactions within a 1D array of spins with spin-1/2 operators $S^\mu$.  Physically, the two spin degrees of freedom correspond to two hyperfine spin states of $^7$Li.  
To probe the dynamics of this model, they initialized the spins in a classical $x$--$z$ helix state, where the spin on site $j$ was oriented with 
$\expv{S^x_j}=\frac{\hbar}{2} \cos(Qj+\phi)$ and $\expv{S^z_j}=\frac{\hbar}{2} \sin(Qj+\phi)$.
The wave vector $Q$ and phase $\phi$ were varied.
The experimentalists quantified the dynamics by studying  the Fourier component of $\langle S^z\rangle$ at wave vector $Q$, a quantity referred to as the {\it contrast}. 
They found that the contrast 
decayed exponentially in time to a non-zero value. From the $Q$ dependence of the decay time, $\tau(Q)$, they identified 
a variety of transport regimes, ranging from ballistic to subdiffusive as a function of anisotropy $J_z$. 

Unfortunately, these observations disagree with theoretical modeling \cite{PereiraSpinHelix2022, VasseurHelixPRB2023}.
The $XX$ limit of the 1D $XXZ$ model (i.e. taking $J_z = 0$) 
can be mapped onto a problem involving free spinless fermions, and hence is exactly solvable.
In Ref.~\cite{PereiraSpinHelix2022} we used this mapping to show that in this $XX$ limit the contrast decays to \textit{zero} as a \textit{power law}, implying that $\tau=\infty$, independent of $Q$. 
Calculations based upon generalized hydrodynamics came to a similar conclusion \cite{VasseurHelixPRB2023}.

Here we consider one possible source of this discrepancy, namely the presence of missing spins. When atoms are loaded into the optical lattice some sites remain empty.   
Aside from a region at the very center of the harmonic trap, these holes cannot move, and hence we treat them as immobile.  To understand the hole mobility, we note that the hopping strength \cite{JepsenNature2020} is $t \approx h\times400$ Hz, and the trap potential is $V_j=(1/2)  \kappa j^2$ with $\kappa \approx h\times100$ Hz.  By energy conservation, a hole at site $j$ can only hop when $V_{j+1}-V_j \approx  \kappa j$ is smaller than $2t$.  Thus, in a chain of length $L\approx 40$ \cite{JepsenNature2020}, only  holes in the central $(2t/\kappa)/L = 20\%$ of the trap can even hop by a single site.  If we extend the criteria of `immobile' to mean that a hole can hop by no more than a single site, then the holes are immobile in more than 90\% of the cloud.  Any disorder in the potential further increases this fraction.
Mobile holes can also be modelled (for example, using the tensor network approaches in Refs.~\cite{JepsenNature2020, Kiely2023}), but given the small region they are contained in, they should make a small impact on the experiment.

It is simple to add immobile holes to the spin model.  The empty sites act as barriers, breaking the spin chain into disjoint segments whose dynamics are independent.  We choose to model the hole density as uniform. 
Calculating the experimental hole distribution is challenging, as it is sensitive to details of how the atoms are loaded into the optical lattice.  Gross features should be captured by a uniform distribution.

The experimentalists estimate that $p\sim5$--$10$\% of the sites are empty \cite{JepsenNature2020}. 
At such small hole densities, we find that the contrast decays exponentially with a time constant $\tau(Q)\propto Q^{-\alpha}$ for $\alpha \sim 1$, which agrees with the experimental observations.  We have quantitative agreement if we take $p=5\%$.

At large hole density ($p\gtrsim 35\%$) we find a distinct dynamical regime, which has not yet been experimentally observed.  
We argue that large hole densities stabilize the helix, preventing its decay.  In this regime we instead observe persistent oscillations of the contrast about a non-zero average. 

Our treatment is exact, but it relies upon special properties of the $XX$ model.  It does not readily generalize to the case where $J_z\neq 0$. 
Thus we also develop a more general matrix product state approach \cite{WhiteDMRGPRL1992} for calculating quantum dynamics in the presence of a random collection of static holes. To perform the average over the distribution of empty sites we borrow a strategy from studies of thermal systems \cite{WhiteTempDMRGPRB2005, SchollwockReview2011}:  we introduce a set of ancillary spins which are entangled with our physical spins.  Tracing over the ancillary spins yields a mixed density matrix for the physical state -- corresponding to an average over all disorder realizations. 
We use the time-dependent variational principle (TDVP) algorithm \cite{YangWhitePRB2020, HaegemanPRB2016} to time evolve this {\it purified density matrix}.
We use our previous modeling to validate this numerical technique, but reserve studies of generic anisotropies for future work.

We emphasize that this study should not be interpreted as a criticism of Ref.~\cite{JepsenNature2020} or other quantum simulators.  Rather, we are in the early days of quantum simulation and therefore must explore and understand the impact of various imperfections.  Similarly, our goal is not to comprehensively model Ref.~\cite{JepsenNature2020} in its particular experimental details.  Our goal is instead to explore the effect of holes on spin dynamics in a tunable manner, exposing the fundamental physics and clarifying why theory and experiment may have disagreed. Confronting and understanding such experiment-theory discrepancies are imperative for developing future generations of simulators.

The outline of this paper is as follows. In Sec. \ref{sec:Setup}, we describe our model and the main observable. In Sec. \ref{sec:Methods}, we 
show how to calculate the properties of this model by considering ensembles of non-interacting fermions.
We present the results of these calculations in Sec. \ref{sec:Results}. In Sec. \ref{sec:ExpComp}, we compare our results to those of the experiment. We conclude with a summary of our work in Sec. \ref{sec:conclusion}.  Appendix~\ref{subsec:DMRGAppendix} describes our matrix product state approach to solving this problem, and Appendix~\ref{subsec:ThermHoleAppendix} describes a rough approximation which relates the thermal fraction of atoms in the experiment to the hole density used in our modeling. 

\section{Setup} 
\label{sec:Setup}

\begin{figure}
    \centering    
    \includegraphics[width=0.8\linewidth]{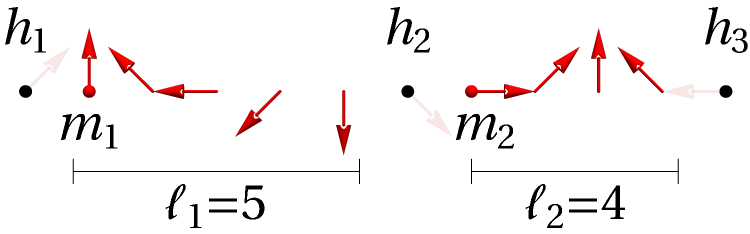}
    \caption{Example configuration of a spin helix with holes. The spin helix has wavelength $\lambda = 8$ and 12 sites are shown. Solid red arrows represent sites with spins; black dots represent the location of immobile holes; transparent red arrows represent the spins that have been replaced by immobile holes; red dots represent starting positions of spinful segments. The 12-site chain in this example has been divided into spinful segments with lengths $\ell_1 = 5$ and $\ell_2 = 4$, starting positions $m_1=2$ and $m_2 = 8$, separated by holes at sites $h_1 = 1, h_2 = 7$ and $h_3 = 12$. The wavefunction for this particular configuration is written $\ket{\psi} = \ket{0}_{h_1} \ket{m_1 \ell_1} \ket{0}_{h_2} \ket{m_2 \ell_2} \ket{0}_{h_3}
    =
    \ket{0}_{1} \ket{2,5} \ket{0}_{7} \ket{8,4} \ket{0}_{12}
    $.}
    \label{fig:SpinChain}
\end{figure}

We take a 1D chain of sites, labeled by integers $j$, that can be in one of three states: $\ket{\uparrow}_j$,
$\ket{\downarrow}_j$,
or $\ket{0}_j$, corresponding to the presence of a spin-$\uparrow$ particle, a spin-$\downarrow$ particle, or an empty site.  These empty sites are treated as immobile, but the spins interact via a $XX$ Hamiltonian
\begin{eqnarray}
    \ham &=& J \sum_j \left[S_j^x S_{j+1}^x + S_j^y S_{j+1}^y\right]  \\
    &=& \frac{J}{2}\sum_j\left[S_j^+S_{j+1}^-+S_j^-S_{j+1}^+\right].
\end{eqnarray}
Here, $S^\mu=\frac{1}{2} \sigma^\mu$ $(\mu = x,y,z)$ are the standard spin-1/2 matrices, with $S^{\pm} = S^x \pm iS^y$. These can be extended into our larger local Hilbert space by taking them to vanish when acting on $\ket{0}$.  We use units where $\hbar=1$.

To model the experiment, we consider an ensemble of initial product states, each of which has the form  $\ket{\psi}=\prod_j \ket{\psi}_j$. The wavefunction on site $j$ is $\ket{\psi}_j=\ket{0}_j$ with 
probability $p$, or $\ket{\psi}_j=\left[A_{j,\uparrow} \ket{\uparrow}_j + A_{j,\downarrow}\ket{\downarrow}_j \right]\equiv \ket{\chi}_j$ with probability $(1-p)$. Here $A_{j,\uparrow}= \sin(\theta_j/2) $ and $A_{j,\downarrow}= \cos(\theta_j/2)$ with $\theta_j=Qj+\phi$ for some wave vector $Q= \frac{2\pi}{\lambda}$ and phase $\phi$. The helix wavelength is $\lambda$.  For our numerics we use $\phi=0$, and we use $\lambda=8$ whenever the pitch dependence is not needed.

Our calculations are simplest when we choose $\lambda$ to be an integer.  We expect that all physics quantities are smooth functions of the wavelength, and hence it is justifiable to limit ourselves to considering integer $\lambda$.

Consider a realization where the holes are at locations $\{h_\nu\}$. 
These holes break the chain into disjoint segments.
At all times the wavefunction 
takes on a product form, with each segment being uncorrelated with the others.
We write this product as $|\psi\rangle = \prod_\nu |0\rangle_{h_\nu} \ket{\Psi}_{h_{\nu}+1,h_{\nu+1}-1},$ where  $\ket{\Psi}_{h_{\nu}+1,h_{\nu+1}-1}$ is the wavefunction for all sites in between the holes at $h_{\nu}$ and $h_{\nu+1}$. We find it convenient to write $\ket{\Psi}_{h_{\nu}+1,h_{\nu+1}-1} \equiv |m_\nu\ell_\nu\rangle$, where $m_\nu$ is the first site in the $\nu$-th spinful chain segment and $\ell_\nu$ is the length of the segment; see Fig. \ref{fig:SpinChain}. Explicitly, $m_\nu=h_\nu+1$ and $m_\nu+\ell_\nu-1=h_{\nu+1}-1$.  
That is, $|2,5\rangle$ would represent a wavefunction for spins $2,3,4,5,6$, while $|8,4\rangle$ describes  a wavefunction for spins $8,9,10,11$ -- as illustrated in Fig.~\ref{fig:SpinChain}.
 
At time $t=0$ (and suppressing $\nu$), the wavefunction for a spinful segment is equivalent to the initial spin helix state on that segment, $\ket{m\ell} = \prod_{j=m}^{m+\ell-1} \left[A_{j,\uparrow} \ket{\uparrow}_j + A_{j,\downarrow}\ket{\downarrow}_j \right] = \prod_{j=m}^{m+\ell-1}  \ket{\chi}_j$. 
To quantify the dynamics of this helix, we calculate the {\it contrast}, which is the primary diagnostic in the experiments. The contrast is the spatial Fourier transform of the $z$-component of the spin texture,
\begin{equation}
     C_Q(t,p) =
     \frac{4}{L} \sum_{j=1}^L e^{iQj} \ensavg{\expval{S^z_j(t,p)}}, \label{eq:Contrast}
\end{equation}
where the bar represents an ensemble average and  $\langle\,\cdot\,\rangle$ the quantum mechanical expectation value in a given realization.  Here $t$ is time, and, as already introduced, $p$ is the probability that any given site contains a hole.
The normalization of ${C}_Q(t,p)$ is chosen so that ${C}_Q(0,p)=1-p$, which is unity when $p=0$.  To compare with the experiments, we focus on two quantities of interest: (i) the time series of the contrast (i.e. ${C}_Q(t,p)$ itself) and (ii) the static background contrast (i.e. the zero-frequency contribution to ${C}_Q(t,p)$, which we denote as $\timeft{C}_Q(\omega=0, p) \equiv \timeft{C}_Q(p)$). 

\section{Methods} 
\label{sec:Methods}

The $XX$ model with holes can be mapped onto non-interacting fermions via a Jordan-Wigner transformation \cite{LSMAnnals1961, JordanWigner, PereiraSpinHelix2022}.
In particular, if the site $j$ is not occupied by a hole, $S_j^+ = e^{-i \pi \sum_{l<j} n_l} a_j^\dagger$ and $S_j^z = n_j - \frac{1}{2}$, where $a_j^\dagger$ is the fermion creation operator and $n_j=a_j^\dagger a_j= (2S_j^z+1)/2$
is the fermion number operator. 
Hence, in terms of the occupation number, 
\begin{align}
    {C}_Q(t,p) &=
     \frac{4}{L} \sum_{j=1}^L e^{iQj} \ensavg{\expval{n_j(t,p)}},  \label{eq:ContrastOccNum} \\
     \timeft{C}_Q(p) &= \frac{4}{L} \sum_{j=1}^L e^{iQj} \expval{\timeft{n}_j(\omega=0,p)} \label{eq:ContrastStatic}.
\end{align}
Determining ${C}_Q(t,p)$ or $\widetilde{C}_Q(p)$ thus reduces to determining $\ensavg{\expval{n_j(t,p)}}$ or its time-average $\expval{\timeft{n}_j(\omega=0,p)}=\lim_{T\to\infty}T^{-1}\int_0^T \ensavg{\expval{n_j(t,p)}}\ dt$.
We will first focus on calculating $\ensavg{\expval{n_j(t,p)}}$.

Consider a particular chain segment $\ket{m\ell}$; we suppress the subscript $\nu$. In this case, the bracketing holes are at sites $m-1$ and $m+\ell$, and the fermionic Hamiltonian for the segment $m \leq j \leq m+\ell-1$ is
\begin{equation}
    \ham_{\rm f} = \frac{J}{2} \sum_{j=m}^{m+\ell-1} \left[a_j^\dagger a_{j+1} + a_{j+1}^\dagger a_j\right].
\end{equation}
The density of fermions on site $j$ at time $t$ within this segment is
\begin{equation}
    \expval{n_j(t)} = \bra{m\ell} a_j^\dagger(t) a_j(t) \ket{m\ell}.
\end{equation}
For non-interacting fermions, the annihilation operator at time $t$ can be written in terms of a $\ell\times\ell$ matrix $G$ as
\begin{equation}
    a_j(t) = \sum_u G_{ju}(t) a_u(0),
\end{equation}
where the Green's function $G$ has elements
\begin{eqnarray}
    G_{ju}(t) &=& \sum_{\mu=1}^{\ell} G_{ju}^\mu(t), \\
    G_{ju}^\mu(t) &=& e^{-i\omega_\mu t} (\vect{v}_\mu^*)_j (\vect{v}_\mu)_u. \label{eq:GreenFunctionExpr}
\end{eqnarray}
Here, $\omega_\mu$ and $\vect{v}_\mu$ are the eigenvalues and eigenvectors of the $\ell \times \ell$ tridiagonal matrix with zeroes on the main diagonal and $J/2$ on the others. 
These wavefunctions correspond to discretized ``particle-in-a-box'' solutions and 
only depend on where $j$ and $l$ sit inside the segment, and not on the absolute location of the segment.  Relative to the hole at site $m-1$, we introduce $\bar j=j-(m-1)$ and $\bar u=u-(m-1)$.  Thus $\vect{v}_\mu$ is a sinusoidal function which vanishes at $\bar j=0,\ell+1$,
\begin{equation}
    (\vect{v}_\mu)_{\bar j} = \sqrt{\frac{2}{\ell+1}} \sin(\frac{\mu \pi }{\ell + 1} \bar{j}).
\end{equation}
In terms of these wavefunctions we can then calculate
\begin{equation}
    \expval{n_j(t)} = \sum_{uw} \sum_{\mu, \nu=1}^\ell (G_{uj}^\mu(t))^* G_{jw}^\nu(t)\bra{m\ell} a_u^\dagger(0) a_w(0) \ket{m\ell}. \label{eq:OccNumGF}
\end{equation}
The initial expectation value $\rho_{uw} = \bra{m\ell} a_u^\dagger(0) a_w(0) \ket{m\ell}$  is given by \cite{PereiraSpinHelix2022}
\begin{equation}
    \rho_{uw} = A_{w, \uparrow} A_{w, \downarrow}^* A_{u, \downarrow} A_{u, \uparrow}^*  \left(\prod_{i=u+1}^{w-1} \left(\abs{A_{i, \downarrow}}^2 - \abs{A_{i, \uparrow}}^2\right)\right)
\end{equation}
for $u < w$, $\rho_{uw} = \rho_{wu}$ for $u > w$, and $\rho_{uu} = |A_{u,\uparrow}|^2$ for $u=w$.
Note,  $\ket{m\ell}$ is a segment of length $\ell$ surrounded on either side by holes. The probability of finding a hole, followed by $\ell$ spins, followed by another hole is
$p \times (1-p)^\ell \times p = p^2(1-p)^\ell$. Hence, summing over all such segments, the ensemble average is accomplished by taking
\begin{align}
    \ensavg{\expval{n_j(t,p)}} &= \sum_{\substack{m\leq j\\\ell +m>j }}  p^2 (1-p)^\ell \bra{m\ell} a_j^\dagger(t) a_j(t) \ket{m\ell}
    \label{eq:OccNumEA}.
\end{align}
Here, the limits on the sum ensure site $j$ is contained within $\ket{ml}$.  Equation~\eqref{eq:ContrastOccNum} is then readily calculated as
\begin{align}
    {C}_Q(t,p) &=
     \frac{4}{L} \sum_{j=1}^L \sum_{\substack{m\leq j\\\ell +m>j }}  p^2 (1-p)^\ell e^{iQj} \bra{m\ell} a_j^\dagger(t) a_j(t) \ket{m\ell}
     \label{eq:EATimeContrast}
\end{align}
with $\expval{{n_j(t)}} = \bra{m\ell} a_j^\dagger(t) a_j(t) \ket{m\ell}$ coming from Eq.~\eqref{eq:OccNumGF}.

Equation \eqref{eq:EATimeContrast} can be written in a more practical form with three simplifications. Firstly, the spin helix has periodicity in $\lambda$. This means the sum over sites need not run over the full chain length $L$. Secondly, the triple sum over $m,j,$ and $\ell$ can be rearranged so that $j$ is constrained by the choice of $m$ and $\ell$. Combining these two simplifications gives
\begin{align}
\label{cqsum}
    {C}_Q(t,p) =
     \frac{4}{\lambda} \sum_{\ell=1}^\infty \sum_{m=1}^\lambda \sum_{m-1<j<m+\ell}  &p^2 (1-p)^\ell e^{iQj} \times \nonumber \\ &\bra{m\ell} a_j^\dagger(t) a_j(t) \ket{m\ell}.
\end{align}
Thirdly, $\expval{n_j(t)} = \bra{m\ell} a_j^\dagger(t) a_j(t) \ket{m\ell}$ can be substituted in from Eq.~\eqref{eq:OccNumGF}, with the Green's functions having the known form from Eq.~\eqref{eq:GreenFunctionExpr}. With this substitution, all summations neatly separate into a telescoping set of expressions, which 
are the key equations that are used to numerically calculate our results:
\begin{align}
    {C}_Q(t,p) &= \sum_{\ell=1}^\infty  p^2 (1-p)^\ell C^\ell_Q(t) \label{eq:SimpleEA}, \\
    C^\ell_Q(t) &= \sum_{\mu, \nu=1}^\ell e^{i (\omega_\mu - \omega_\nu) t} C_{\mu\nu}^\ell(Q), \label{eq:SimpleTE} \\
    C_{\mu \nu}^\ell(Q) &= \frac{4}{\lambda} \left[\sum_{m = 0}^{\lambda-1} \sum_{u,w=1}^{\ell} e^{iQm} \rho_{u+m,w+m} (\vect{v}_\mu)_u^* (\vect{v}_\nu)_w \right] \times \nonumber \\ &\qquad \left[\sum_{j=1}^{\ell}  e^{iQj} (\vect{v}_\mu)_j^* (\vect{v}_\nu)_j\right] \label{eq:SimpleMoments}.
\end{align}
The summation over $\ell$ in Eq.~\eqref{eq:SimpleEA} performs the ensemble average for a given hole probability $p$. The summation over momenta $\mu$ and $\nu$ in Eq.~\eqref{eq:SimpleTE} produces the time evolution. The moments $C_{\mu \nu}^\ell(Q)$ in Eq.~\eqref{eq:SimpleMoments} contain all 
information about the spin helix.  It is the only sum involving the site indices.  The sum over $j$ can be performed analytically, by writing the sinusoidal wavefunctions in terms of complex exponentials and evaluating the resulting geometric series. Thus tabulating the $C_Q^\ell$'s for a single $t$ takes $\mathcal{O}(\lambda\ell^4)$ operations.  These coefficients are independent of $p$, and then can be summed in Eq.~(\ref{eq:SimpleEA}) to arrive at the time-dependent contrast for arbitrary $p$.

\begin{figure*}[t!]
    \centering    
    \includegraphics[width=\linewidth]{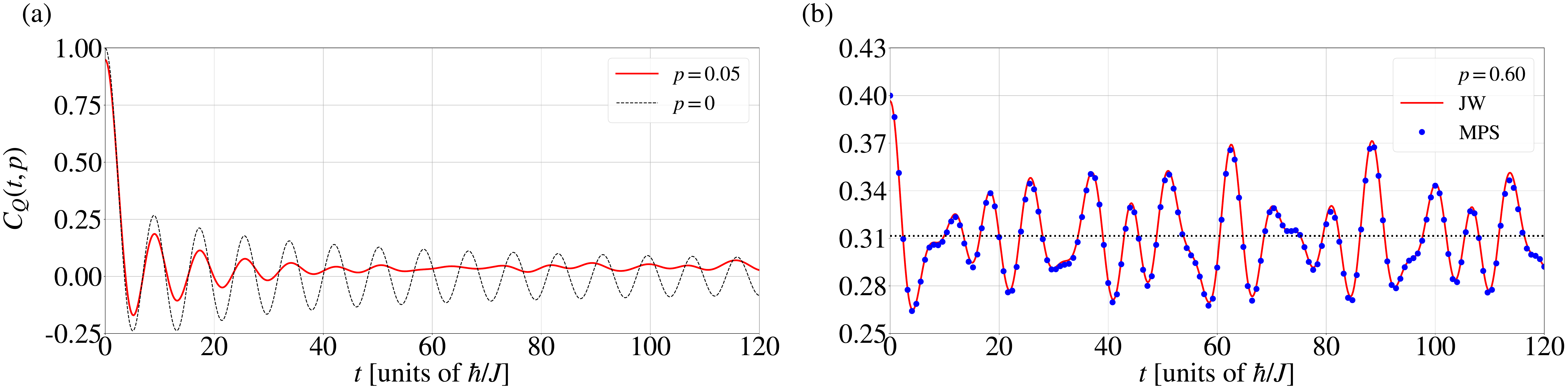}
    \caption{Ensemble-averaged contrast ${C}_Q(t,p)$ for helix wavelength $\lambda=8$, phase $\phi = 0$,  in three distinct dynamical regimes. (a) ${C}_Q(t,p=0.05)$ calculated by mapping onto non-interacting fermions using a Jordan-Wigner transformation (solid red line). It exhibits roughly exponential decay at early times. The exact result in the absence of holes \cite{PereiraSpinHelix2022} is also displayed (dashed black line), exhibiting power law behavior. (b) ${C}_Q(t,p=0.60)$ calculated using the Jordan-Wigner (JW) method described in the main text (solid red line) as well as the numerical Matrix Product State (MPS) method described in Appendix \ref{subsec:DMRGAppendix} (blue circles).  In this regime one sees  persistent oscillations. The dotted black line shows $\timeft{C}_Q(p=0.6)$.}
    \label{fig:TimeSeries}
\end{figure*}

The static background contrast $\widetilde{C}_Q(p)$ is readily calculated from these expressions. Equation \eqref{eq:SimpleTE} controls the time evolution of the spin helix. The particle-in-a-box spectrum $\{\omega_\mu\}$ is non-degenerate for any choice of $\ell$. Hence, the zero-frequency contribution comes only from those terms in Eq.~\eqref{eq:SimpleTE} for which $\mu=\nu$:
\begin{align}
    \widetilde{C}_Q(p) &= \sum_{\ell=1}^\infty  p^2 (1-p)^\ell C^\ell_Q(\omega=0), \\
    C^\ell_Q(\omega=0) &= \sum_{\mu=1}^\ell C_{\mu\mu}^\ell(Q).
\end{align}
These can be efficiently calculated.

The probability that a spin is in a segment longer than $\ell_{\rm max}$ sites is
\begin{equation}
    P_{\ell>\ell_{\rm max}}= (1-p)^{\ell_{\rm max} + 1}(1+p\ell_{\rm max}). \label{eq:cumulative}
\end{equation}
To numerically evaluate Eq.~\eqref{eq:SimpleEA} we introduce a cutoff $\ell_{\rm max}=300$ such that $P_{\ell>\ell_{\rm max}}\lesssim10^{-3}$ for the smallest $p=0.03$ that we consider.  This $\ell_{\rm max}$ is larger than the experimental system size of $\sim$40 sites.  We use this larger cutoff to be able to model smaller values of $p$.  To prove Eq.~\eqref{eq:cumulative}, we note that there are $\ell$ possible segments of length $\ell$ that can contain a given spin. Such segments occur with probability $p^2 (1-p)^\ell$, and hence $P_{\ell>\ell_{\rm max}}=\sum_{\ell=\ell_{\rm max}+1}^\infty \ell p^2 (1-p)^\ell$, which evaluates to Eq.~\eqref{eq:cumulative}.

Although we assume an integer $\lambda$, this calculation can readily be extended to 
rational wavelength $\lambda=r/q$. The primary change would be that the sum over $m$ in Eq.~(\ref{cqsum}) would instead run from 1 to $(q\times\lambda)$.  An irrational $\lambda$ can be modelled as the limit of a series of rational approximants.  As already argued, we expect the experimental observables to be smooth functions of $\lambda$, and the results with integer $\lambda$ should be representative.

\section{Results}
\label{sec:Results}

Figure \ref{fig:TimeSeries} shows the ensemble-averaged contrast ${C}_Q(t,p)$ for $\lambda = 8$, $\phi = 0$, and three different choices of the hole probability: $p=0$ and $p=0.05$ in Fig.~\ref{fig:TimeSeries}\red{(a)} and $p=0.60$ in Fig.~\ref{fig:TimeSeries}\red{(b)}. These parameters illustrate the three regimes that we observe:
(1) In the absence of holes, $p=0$, the contrast oscillates with an envelope that falls off as a power law, $C(t) \sim t^{-1/2}$.  At long times $C(t)$ approaches zero.
(2) At small but non-zero $p$, we initially see an exponential-like envelope (until $t \approx 50 \hbar/J$ for $p=0.05$), followed by weak but long-lived oscillations about a non-zero mean. This is the regime most relevant to experiments.  (3) At large $p$ we see large aperiodic oscillations about a non-zero mean.  In Fig.~\ref{fig:TimeSeries}\red{(b)}, we also show the results of our numerical matrix product state calculation (described in Appendix \ref{subsec:DMRGAppendix}).  They are indistinguishable from our Jordan-Wigner approach. 

In the context of Eqs.~\eqref{eq:SimpleEA}--\eqref{eq:SimpleMoments}, the exponential-like decay in Fig.~\ref{fig:TimeSeries}\red{(a)} can be understood from the segment-length dependence of the fermion spectrum in the low-$p$ regime. Each segment length $\ell$ contributes a different set of frequencies and hence a different set of phase factors in the time evolution, Eq.~\eqref{eq:SimpleTE}, leading to dephasing. At low $p$, the segment length distribution is quite broad,
producing a decay whose time constant will grow with decreasing $p$ (an observation to be discussed later on in this section). 
At large $p$ most segments are very short, resulting instead in only a few discrete frequencies. These are incommensurate with one another, leading to persistent quasiperiodic oscillations.  The long-time behavior at small $p$ is similar; there are just more discrete frequencies involved, and hence the oscillations are weaker.

Perhaps the most notable feature in these graphs is the long-time non-zero background contrast $\widetilde{C}_Q(p)$.  This is best understood by noting that the net spin polarization is conserved in any given segment:  a segment which initially has a large total $\langle S^z\rangle$ will always have a large net polarization;  a segment with small total $\langle S^z\rangle$ will always have a small net polarization. Since all segments are separated by holes and cannot equilibrate with each other, some memory of the spatial spin patterns persists for all times. We expect this background to tend to zero as $p \to 0$ (where equilibration occurs across large portions of the chain) and as $p \to 1$ (where most of the initial polarization is lost to holes).

\begin{figure}[t!]
    \centering    
    \includegraphics[width=\linewidth]{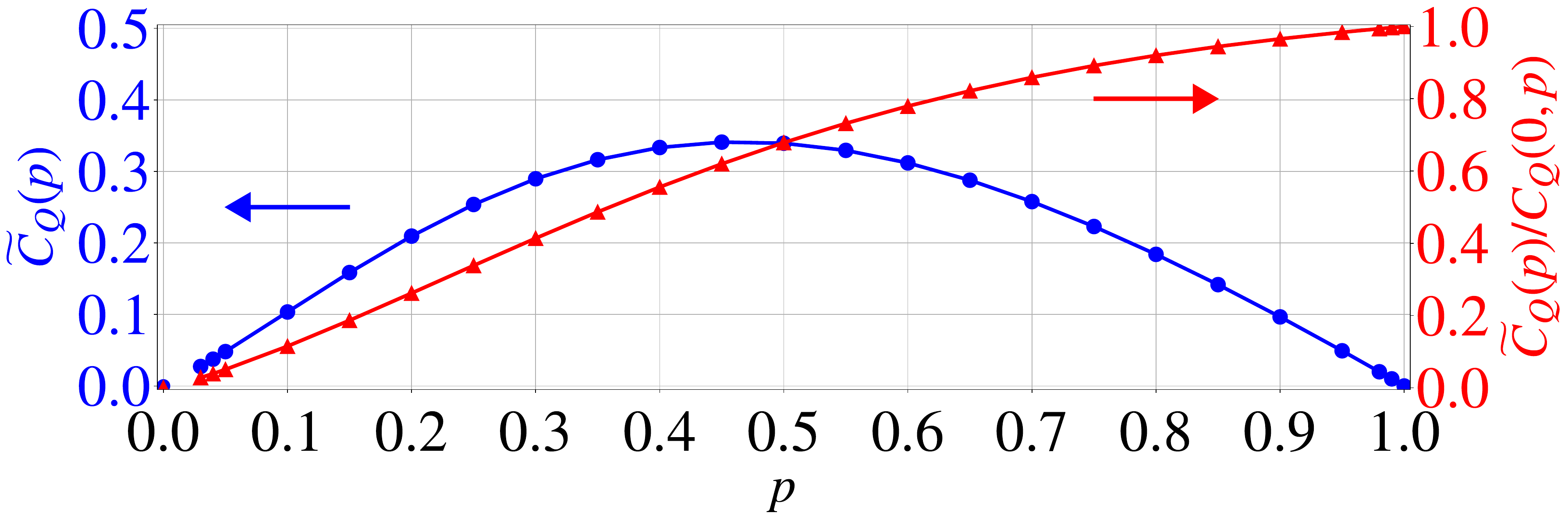}
    \caption{Static background contrast $\timeft{C}_Q(p)$ (blue circles, left axis) as a function of hole probability $p$ for helix wavelength $\lambda = 8$ and phase $\phi = 0$. $\timeft{C}_Q(p)$ normalized by the initial contrast ${C}_Q(0,p) = 1-p$ is also shown (red triangles, right axis). The data points at $p=0$ are known from previous work \cite{PereiraSpinHelix2022}.
    }
    \label{fig:LongTime}
\end{figure}

This static background contrast $\widetilde{C}_Q(p)$ is shown in Fig.~\ref{fig:LongTime} as a function of $p$. Indeed, the background contrast tends to zero as $p \to 0$ and $p \to 1$. 
We also show the ratio of the static background contrast to the initial contrast, ${C}_Q(0,p) = 1-p$, corresponding to the fraction of the initial contrast that remains at long time.  This normalized contrast is a monotonically increasing function of $p$. 

For $p\lesssim 0.35$ we observe a notable separation between a rapidly decaying envelope at short times and aperiodic oscillations at long times.  For larger $p$ the decay time is so short that one cannot reliably make such a separation.  To extract the decay rate at small $p$ ($\lesssim 0.2$), we 
fit the {\it envelope} of the contrast versus time curves to a function of the form
\begin{equation} 
\label{eq:CrossoverFit}
C_Q^{\rm env}(t,p)= \widetilde C_Q(p)+A t^{-1/2} e^{-\Gamma t}.
\end{equation}

We shift by 
$\widetilde{C}_Q(p)$ to account for the background contrast.
The factor of $t^{-1/2}$ is included so that the envelope has the correct functional form when $p=0$.
For $\lambda=8$, we empirically find $\widetilde{C}_Q(p) \approx p$ for small $p$ 
(see Fig.~\ref{fig:LongTime}).
To find the optimal $A$ and $\Gamma$, we perform a least-squares fit 
for $0<t \lesssim 40\hbar/J$, which excludes any of the long-time persistent oscillations for $p \in [0.03, 0.09]$. The resulting decay rate $\Gamma$ as a function of $p$ is shown in Fig.~\ref{fig:ExpPowCrossover}. When $p$ is finite, the decay rate $\Gamma$ is non-negligible, corresponding to an exponential decay (with logarithmic corrections). As $p \to 0$, however, $\Gamma \to 0$, indicating a diverging time constant with decreasing $p$. Ultimately, an exponential decay at small $p$ gives way to a power law decay as $p \to 0$. This offers one possible resolution to the discrepancy between earlier hole-free calculations and the experiments.

\section{Comparison to Experiment and Prior Modeling}
\label{sec:ExpComp}

By modifying their loading procedure,
Jepsen et al. \cite{JepsenNature2020}
were able to increase their hole density and experimentally study some of its impact.  They analyzed the time series of the contrast, fitting it to an empirical form 
\begin{equation}\label{cemp}
    C_Q^{\rm emp}(t) = \left[a_0 + b_0\cos(\omega t)\right]e^{-t/\tau} + c_0.
\end{equation}
Their fits used data with time between $0$ and $20 \hbar/J$ at large $Q$ and between $0$ and $30 \hbar/J$ at small $Q$.
They  found that in the $XX$ limit the oscillation period of the contrast and the decay time only weakly depended on $p$, with the decay time decreasing slightly as $p$ increased.

On the other hand, the normalized background contrast $c_0$ increased monotonically with hole concentrations, over the range that they explored.  For all $p$
they found that the contrast decayed exponentially to its background value. 

As in our model, the normalized experimental background contrast increased with hole probability, but a quantitative comparison is challenging as they do not have a direct measure of the hole density.  In Appendix~\ref{subsec:ThermHoleAppendix} we use entropy arguments to model the experimental hole density, and make some comparisons.

As already discussed in Sec.~\ref{sec:Results}, at larger hole densities our model displays persistent quasiperiodic oscillations in the contrast.  The experiments do not see these oscillations.  Instead their data is well described by Eq.~(\ref{cemp}), with perhaps an additional long-time drift. 
Clearly there is additional physics needed to explain these large $p$ results.  In the experiment, the harmonic trap leads to an inhomogeneous hole distribution.  There are also mobile holes in the central region of the trap.  Both of these effects are potential sources of the discrepancy.  Nonetheless, it appears that static holes play an important role in the experiment, especially at low hole densities.

\begin{figure}[t!]
    \centering    
    \includegraphics[width=1\linewidth]{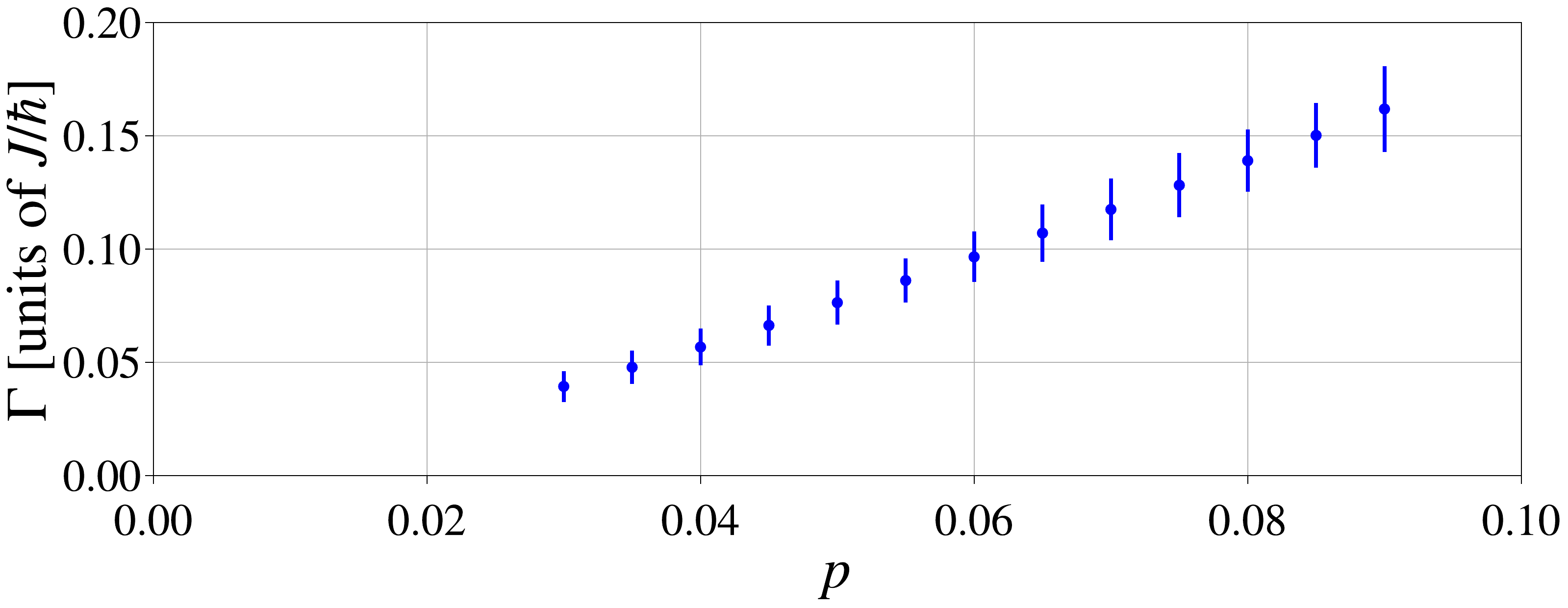}
    \caption{Exponential decay rate $\Gamma$ from Eq.~\eqref{eq:CrossoverFit}, determined by fitting the maxima of the scaled-and-shifted contrast $({C}_Q(t,p)-\widetilde{C}_Q(p))\sqrt{t}$ for various hole probabilities $p$. $\Gamma \to 0$ as $p \to 0$ is indicative of the evolution from exponential decay at nonzero $p$ to power law decay at $p=0$.}
    \label{fig:ExpPowCrossover}
\end{figure}

In Ref.~\cite{JepsenNature2020}, the experimentalists also made comparisons to numerical simulations of the bosonic $t$--$J$ model with finite hole concentrations.  Their simulations did not, however, include the harmonic trap, and hence their holes were mobile throughout rather than fixed.  
Thus their model was very different from ours.
In the 
$XX$ limit, those simulations showed an exponential decay of the contrast.  Crucially, however, they found that the contrast vanished at long times -- that is, the simulations had $\widetilde C_Q(p)=0$. It therefore seems likely that immobile holes are not simply important but even \textit{necessary} for producing the finite background contrast (see Sec.~\ref{sec:Results} for a physical argument). Indeed, the authors of Ref.~\cite{JepsenNature2020} also argued that immobile holes were the source of the finite background contrast.

\begin{figure}[t!]
    \centering    
    \includegraphics[width=0.95\linewidth]{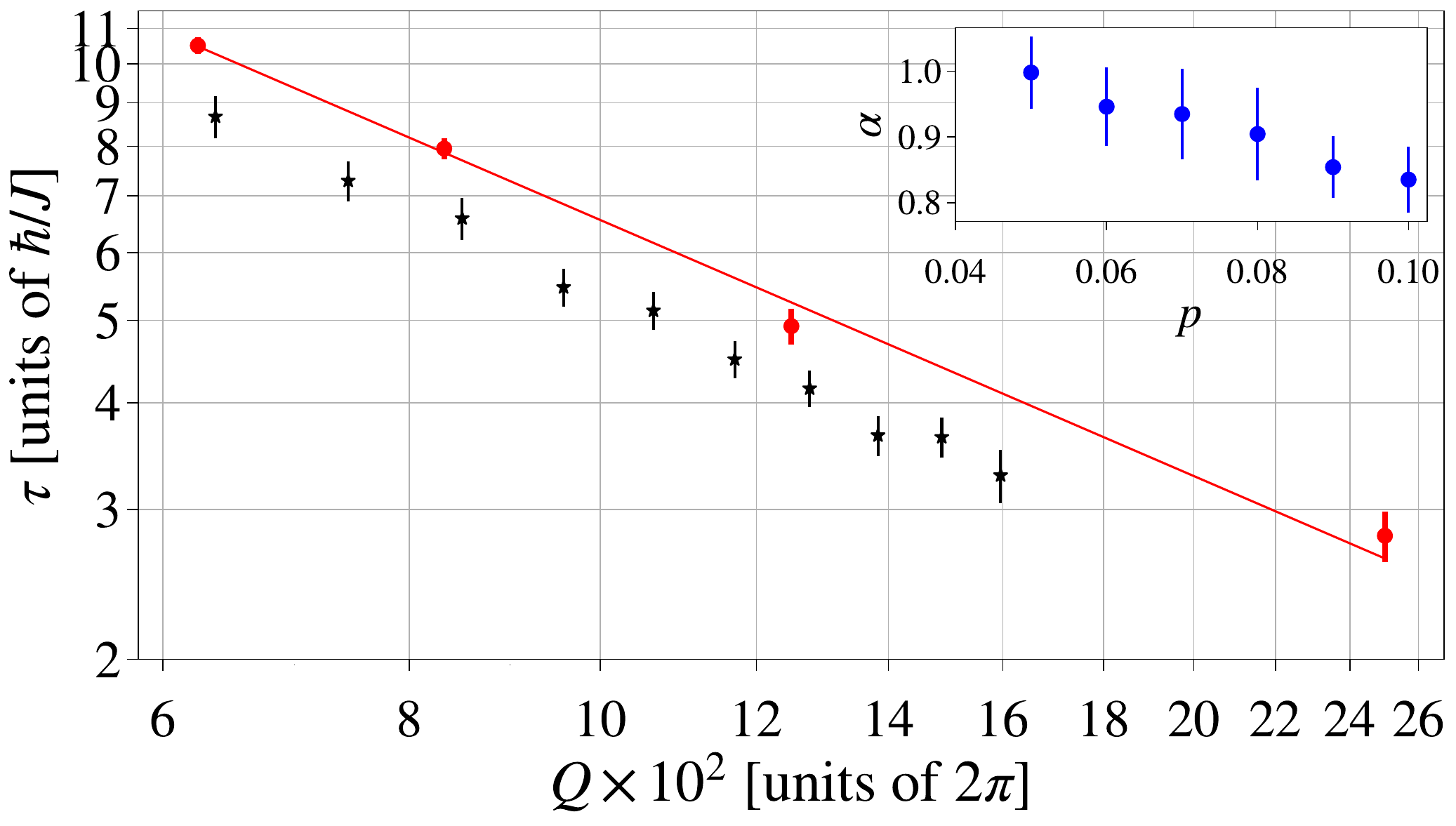}
    \caption{Decay time $\tau$ as a function of helix wave vector $Q$ at $p=0.05$ (red circles), using the same fitting function Eq.~\eqref{cemp} for the contrast as Ref.~\cite{JepsenNature2020}. The straight line represents the least-squares fit of our calculated $\tau(Q)$, with the decay time satisfying $\tau \propto Q^{-\alpha}$ for $\alpha = 1.00(5)$. The black stars are experimental data points from Ref.~\cite{JepsenNature2020}. Inset: decay exponent $\alpha$ as a function of $p$ using the same fitting procedure.}
    \label{fig:ExpDecayTime} 
\end{figure}

In the experiment a central role was played by the 
pitch dependence of the decay time $\tau(Q)$, as they used its behavior (particularly its power law scaling $\tau \sim Q^{-\alpha}$) to distinguish between various transport regimes.
To compare with those results, we use their fitting function Eq.~\eqref{cemp} for the contrast, 
extracting the constants $a_0, b_0, c_0$ and $\tau$
for $\lambda =2\pi/Q= 4, 8, 12, 16$ and $p \in [0.03, 0.10]$. 
We fit the data from time $t=0$ to the time at which quasiperiodic oscillations begin, which ranges from as small as $t\approx 20\hbar/J$ to as large as $t \approx 80 \hbar/J$ depending on $\lambda$ and $p$. In this manner we only fit the regime of exponential decay.
Figure~\ref{fig:ExpDecayTime} shows the resulting best-fit decay time $\tau$ as a function of the wave vector $Q$ on a log-log plot at $p = 0.05$.  The best fit describes a straight line, and from it we conclude that
$\tau(Q) \propto Q^{-\alpha}$ 
with
$\alpha = 1.00(5)$ when $p = 0.05$.  This quantitatively agrees with the exponent measured by the experimentalists. As seen in the inset, our exponent $\alpha$ decreases with increasing hole density, going as low as $\alpha = 0.84(6)$ at $p=0.10$. 

Figure~\ref{fig:ExpDecayTime} also shows the experimental data.  Our predicted decay constants are roughly $20\%$ greater than what was measured in the experiment, but as already emphasized the exponents agree.  This small discrepancy could be due to details in the fitting procedure (for example, the range of times used), or physics which was not included in our model (mobile holes, inhomogeneous hole distribution, etc.).

We caution that the residuals of our fit to Eq.~(\ref{cemp}) are only small at very short times, including no more than two oscillations.  Consequently, the time constant $\tau$ extracted from this fit is not the inverse of the decay rate $\Gamma$ extracted from the envelope, which was plotted in Fig.~\ref{fig:ExpPowCrossover}.

\section{Conclusion}
\label{sec:conclusion}

We have studied the effect of immobile holes on the quantum dynamics of the $XX$ spin helix, revealing three dynamical regimes as a function of hole probability $p$. For $p=0$, the contrast of the spin helix decays to zero as a power law; for small $p$, the contrast decays exponentially to a finite background value, about which it exhibits weak oscillations; for large $p$, the contrast exhibits large quasiperiodic oscillations about a finite background. The experiment largely operated in the regime of small $p$.  We are able to explain a number of their observations, including the finite background contrast and exponential decay of the contrast. We find quantitative agreement with the pitch dependence of their exponential decay constant. As such, a small density of immobile holes is sufficient for explaining the main experimental mysteries.

We caution, however, that in our attempt to produce the simplest and most intuitive picture we have neglected a number of experimental details.  The experiments are performed on an array of finite length spin chains, which may not be identical:  they each contain different numbers of particles, have different hole distributions, and due to field gradients may experience slightly different Hamiltonians.  The particles in each spin chain feel a harmonic potential.  This localizes the majority of the holes, but it also leads to an inhomogeneous hole distribution, with more holes in the wings.  Our modeling also does not take into account the mobile holes which are found in the center of the trap.  All of these effects could be included in finite-chain $t$--$J$ model calculations, at the cost of making the results harder to interpret.
One could also envision modifications of the experiment which would eliminate some of these complications.  For example, adding a large field gradient could ensure that all holes are immobile \cite{KetterlePRL2020}.

Our work serves as a warning for transport studies using analog quantum simulators \cite{JepsenPRX2021, JepsenNatPhys2022}. The spin dynamics for hole probabilities as small as 5\% already significantly differed from the true spin dynamics of the $XX$ spin helix in the absence of holes (see Fig.~\ref{fig:TimeSeries}\red{(a)}). In the experiment, this density of holes on a chain of length $L=40$ corresponds to only two holes. Furthermore, our modeling suggests that the experimental timescales may be  too short to reliably distinguish between an exponential and power law decay.

On a positive note, our study illustrates the richness of the physics which is being explored by the current generation of quantum simulators.  The experiments in Ref.~\cite{JepsenNature2020} have  taught us much about the dynamical behavior of spin chains, and the way that cold atoms can be used to explore that physics.

\section*{acknowledgement}
D. P. is grateful to Thomas G. Kiely for useful discussions. We are also grateful to Wolfgang Ketterle, Paul Niklas Jepsen, Vitaly Fedoseev, Eunice (Yoo Kyung) Lee, Hanzhen Lin, and Andrew Winnicki for sharing their experimental data with us and for comments on our manuscript. This material is based upon work supported by the National Science Foundation under Grant No.  PHY-2110250. We also acknowledge the support of the Natural Sciences and Engineering Research Council of Canada (NSERC) (Ref. No. PGSD-567963-2022).


\appendix
\section{Matrix Product State Method for Hole Ensembles}
\label{subsec:DMRGAppendix}

The properties of an ensemble of quantum states is captured by its density matrix,
\begin{equation}\label{dm}
    \rho_S = \sum_{\alpha=1}^r p_\alpha \ket{\alpha}_S \bra{\alpha}_S.
\end{equation}
Here the probability of finding state $|\alpha\rangle_S$ is $p_\alpha$.  
Note, the set of $\{|\alpha\rangle_S\}$ need not be orthogonal (though in most formulations they are).
It is often helpful to encode this density matrix in a purified wavefunction \cite{WhiteTempDMRGPRB2005, SchollwockReview2011}, defined by
\begin{equation}
\ket{\psi} = \sum_{\alpha=1}^r s_{\alpha} \ket{\alpha}_S  \ket{\alpha}_T. \label{eq:STPureState}
\end{equation}
Here $\{|\alpha\rangle_T\}$ is a set of orthonormal states in some unphysical auxiliary space $T$, and $|s_\alpha|^2=p_\alpha$.  The original density matrix can then be recovered by performing a partial trace 
\begin{equation}
\rho_S = \Tr_T\ket{\psi}\bra{\psi}.
\end{equation}
The construction of the purified wavefunction $|\psi\rangle$ is clearly not unique, as one has a choice of the decomposition in Eq.~(\ref{dm}), the auxilliary space $T$, the vectors $|\alpha\rangle_T$, and the phases of the coefficients $s_\alpha$.  

Such purified wavefunctions are routinely used in matrix product state (MPS) simulations of thermal ensembles  \cite{WhiteTempDMRGPRB2005, SchollwockReview2011}.  Here we use this approach to model the dynamics of an ensemble of spin chains with a random distribution of immobile holes.
As shown in Fig.~\ref{fig:TimeSeries}\red{(b)}, the resulting MPS simulation agrees with our Jordan-Wigner approach to modeling the experiment.  It is more numerically expensive, but unlike our Jordan-Wigner approach can be extended beyond the $XX$ limit.

We describe our purified state as an alternating array of physical and auxiliary sites.  As explained in the main text, the physical site at integer location $j$ can be in one of three states: $\ket{\uparrow}_j,\ket{\downarrow}_j, \ket{0}_j$.  The auxiliary sites  at half-integer positions can be in one of two states: $|0\rangle_{j+1/2},|1\rangle_{j+1/2}$.
Our initial ensemble can then be encoded in a purified wavefunction,
\begin{align}
    \ket{\psi} = \sum_{s}\prod_j &(1-p)^{s_j/2} p^{(1-s_j)/2} \ \times \nonumber \\ &\left[({1-s_j}) \ket{0}_j\ket{0}_{j+1/2}
+ {s_j} \ket{\chi}_j \ket{1}_{j+1/2}\right], \label{eq:HolePsi}
\end{align}
where $s_j=0,1$. The probability of finding a hole on a given site is $p$, and $|\chi\rangle_j$ is given in Sec.~\ref{sec:Setup}.
One can readily verify that
\begin{align}
    \rho(t = 0, p) &= \rho_S = \Tr_T\ket{\psi} \bra{\psi} \\
    &= \prod_j \left[ p \ket{0}_j \bra{0}_j + (1-p) \ket{\chi}_j \bra{\chi}_j  \right], \label{eq:PurifiedHole}
\end{align}
which is the density matrix describing our spin helix after randomly adding immobile holes.

We time-evolve Eq.~(\ref{eq:PurifiedHole}) by using the time dependent variational principle (TDVP) algorithm \cite{HaegemanPRB2016, YangWhitePRB2020} as implemented in the 
ITensor library \cite{ITensorLib}.
We evolve the system from $t = 0$ to $t = 128 \hbar/J$ with time steps of size $\Delta t = 0.8 \hbar/J$. The Hamiltonian only acts on the physical space.  At every time step, we first perform a global subspace expansion \cite{YangWhitePRB2020}, using two Krylov states:
that is, we construct a MPS representation of both 
$\ket{\psi(t)}$ and $\ham \ket{\psi(t)}$, and use the resulting tensors to represent $\ket{\psi(t+\delta t)}$.
For the time evolution using the time-dependent variational principle, we use a singular value decomposition (SVD) cutoff of $10^{-7}$ and a maximum bond dimension of 600, with one sweep being performed at every time step. 
We use a larger cutoff, $10^{-3}$, in our global subspace expansion.

\section{Modeling the Experimental Hole Density} \label{subsec:ThermHoleAppendix}

\begin{figure}[t!]
    \centering    
    \includegraphics[width=0.8\linewidth]{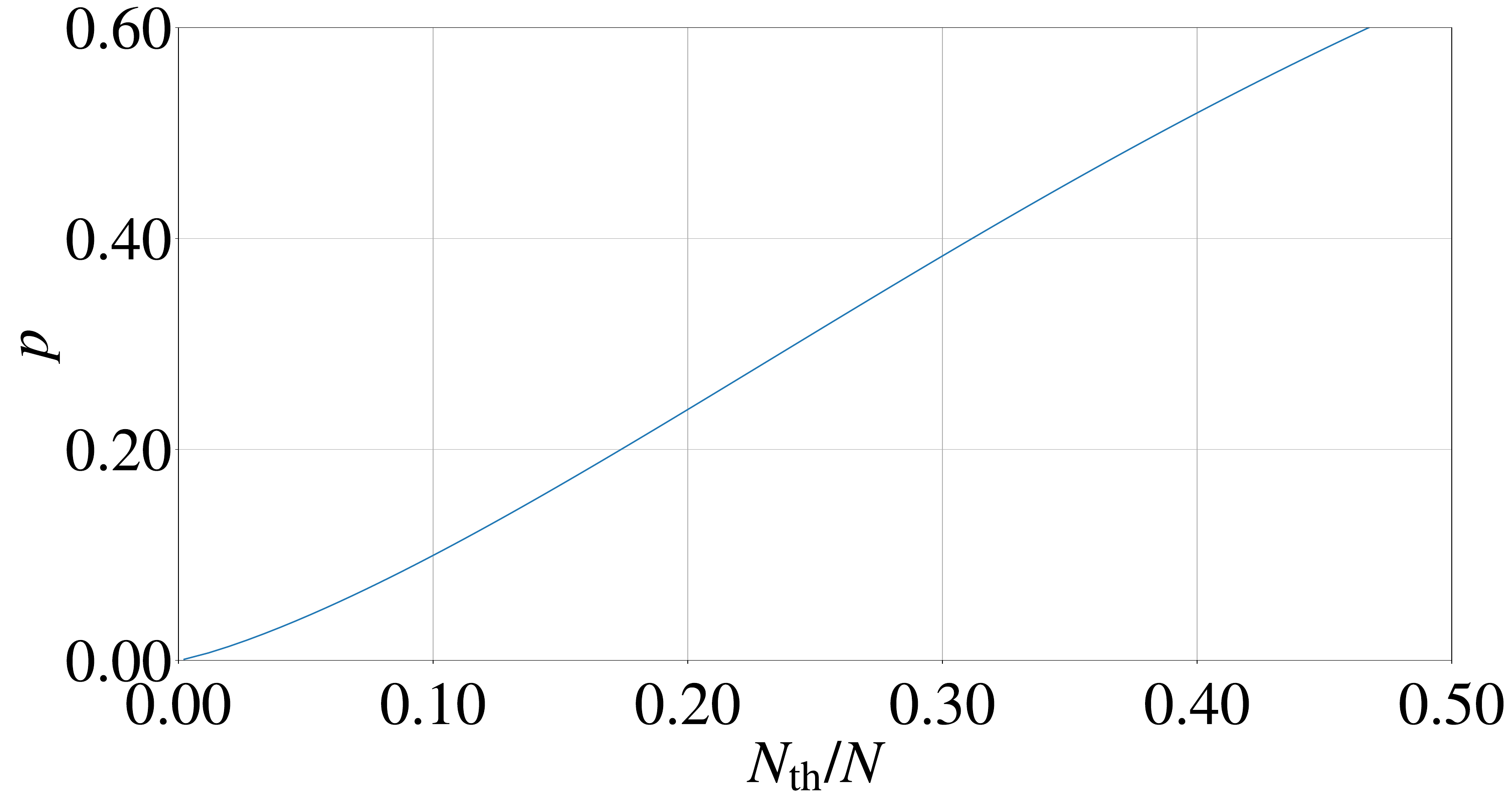}
    \caption{Approximate relationship between hole density $p$ of the Mott insulator, and the thermal fraction $N_{\rm th}/N$ of the harmonically-trapped gas from which it is loaded.  This crude estimate comes from equating the entropies in Eqs.~(\ref{gasS}) and (\ref{holeS}).}
    \label{fig:Entropy}
\end{figure}

\begin{figure}[t!]
    \centering    
    \includegraphics[width=\linewidth]{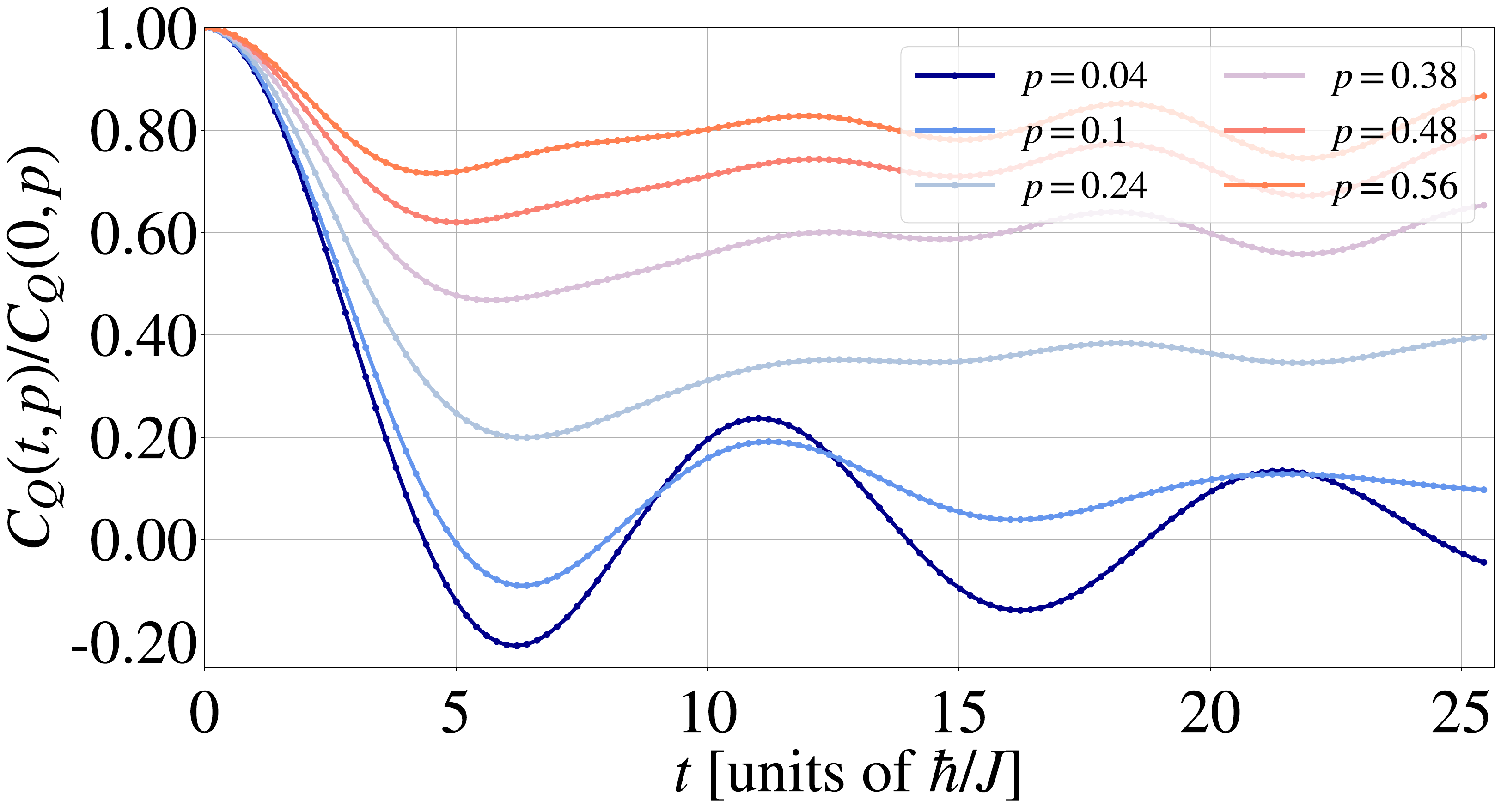}
    \caption{
    Time evolution of the normalized contrast $C_Q(t,p)/C_Q(0,p)$ for a chain with $\lambda=10$, $\phi=0$, and $p=0.04,0.10,0.24,0.38,0.48,$ and $0.56$, which have the same entropy as a harmonically-trapped condensate with thermal fraction 
    $N_{\rm th}/N=0.05,0.1,0.2,0.3,0.37,$ and $0.43$.
    }
    \label{fig:nth}
\end{figure}

In Ref.~\cite{JepsenNature2020}, the experimentalists loaded a three-dimensional optical lattice (consisting of a two-dimensional array of one-dimensional chains) from a harmonically-trapped gas.  They controlled the density of holes by adjusting the temperature of the initial cloud.  A higher temperature cloud has more entropy, resulting in spin chains with more holes.  Here we estimate the hole density by simply equating the entropy of the initial cloud to the configurational entropy from a distribution of holes in a perfect Mott insulator.  

Of course, this is at best a crude approximation.  The loading process will undoubtedly lead to a non-uniform hole density.  Further, these entropy arguments neglect processes which could either increase or decrease the entropy.  Non-adiabatic transitions during the loading process will increase the entropy.  Conversely,  during loading entropy tends to be pushed from the central Mott-insulating region into the superfluid wings.  This leads to an effective entropy reduction in the central region.  It is also worth noting that the loading process generically leads to a non-equilibrium state. Nevertheless, our approach serves as a first-order approximation for estimating the hole density in the experiment.

\begin{figure}[t!]
    \centering    
    \includegraphics[width=\linewidth]{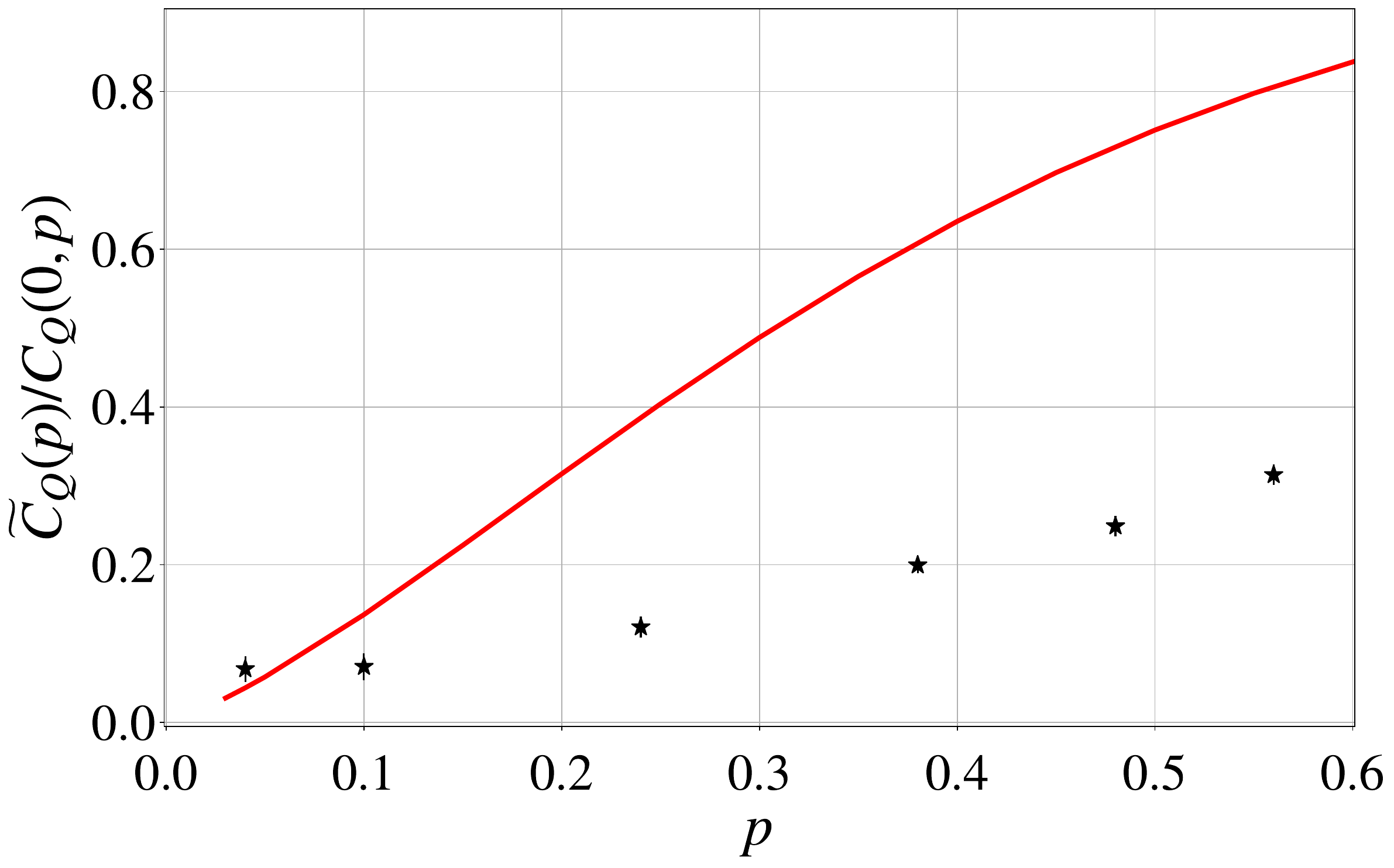}
    \caption{Static background contrast $\timeft{C}_Q(p)$ normalized by the initial contrast ${C}_Q(0,p) = 1-p$ for $\lambda=10$ and $\phi=0$ (solid red line). The black stars are experimental data points from Ref.~\cite{JepsenNature2020} for $\lambda=10.4$ and $p=0.04,0.10,0.24,0.38,0.48,$ and $0.56$, which have the same entropy as a harmonically-trapped condensate with thermal fraction 
    $N_{\rm th}/N=0.05,0.1,0.2,0.3,0.37,$ and $0.43$.}
    \label{fig:CompBack}
\end{figure}

The grand canonical free energy of a three-dimensional harmonically-trapped Bose condensate is 
$
\Omega=-k_{\rm B} T \left({k_{\rm B} T}/{\hbar\omega}\right)^3\zeta(4),
$
where $\zeta(j)=\sum_{s=1}^{\infty} s^{-j}$ is the Riemann zeta function \cite{PhysRevA.35.4354}.  The entropy is $S=-\partial \Omega/\partial  T=4\Omega/ T$.  The number of non-condensed atoms is $N_{\rm th}=(k_{\rm B} T/\hbar \omega)^3 \zeta(3)$, and hence the entropy per particle is
\begin{equation}\label{gasS}
\left(\frac{S}{N}\right)_{\rm gas}=4k_{\rm B}\frac{\zeta(4)}{\zeta(3)} \frac{N_{\rm th}}{N}
\end{equation}
for $N$ particles.

On the other hand, for a random distribution of holes, the entropy is $S=-k_{\rm B} N_s [p\ln p+(1-p)\ln(1-p)]$. Here, $N_s$ is the number of sites and $p$ is the probability that a hole is found on any site.  The number of particles is $N=N_s (1-p)$, and hence the entropy per particle is
\begin{equation}\label{holeS}
\left(\frac{S}{N}\right)_{\rm lattice}=-k_{\rm B}\left(\frac{p}{1-p}\ln p+\ln(1-p)\right)
\end{equation}
Equating the two entropies, Eqs.~\eqref{gasS} and \eqref{holeS}, relates the thermal fraction $N_{\rm th}/N$ of the harmonically-trapped gas to the hole density $p$ in the Mott-insulating chains.
The resulting relation is illustrated in Fig.~\ref{fig:Entropy}.

With this relation in hand, we show the time-dependence of the contrast for various values of $p$ in Fig~\ref{fig:nth}, roughly corresponding to the values of $N_{\rm th}/N$ used in the experiment.  Here we use $\lambda=10$, to allow comparison to the experimental studies with $\lambda=10.4$ \cite{JepsenNature2020}. 
Despite notable quantitative differences, this crude model appears to capture the general trends seen in the equivalent plot in the experimental paper.  In particular, increasing $p$ or $N_{\rm th}$ results in a larger normalized contrast, with weaker oscillations.

Additionally, as shown in Fig.~\ref{fig:CompBack},
 we can use the relationship between $p$ and $N_{\rm th}$ to compare 
the experimental background contrast for various values of $N_{\rm th}/N$ with our calculated background contrast
as a function of the corresponding hole density $p$.  Here we again use $\lambda=10$.  It is clear that for $p>0.05$ our model significantly overestimates the magnitude of the background contrast.  It is difficult to determine the extent to which the discrepancy is due to our estimation of $p$, as opposed to physics that we did not include in our model of the spin dynamics (mobile holes, the harmonic trapping potential, field inhomogeneities, etc.).

%

\end{document}